\documentclass[epj]{svjour}
\usepackage{graphics}
\usepackage[english]{babel}
\usepackage[utf8x]{inputenc}
\usepackage[colorinlistoftodos]{todonotes}
\usepackage[per-mode=symbol]{siunitx}
\usepackage{amsmath}
\usepackage{amssymb}
\usepackage[numbers,sort&compress]{natbib}

\usepackage{graphicx}% Include figure files
\usepackage{dcolumn}% Align table columns on decimal point
\usepackage{bm}% bold math
\newcommand{\de}{\mathrm{d}}
\newcolumntype{P}[1]{>{\centering\arraybackslash}p{#1}}
\begin{document}
\title{Effects of the second virial coefficient on the adiabatic lapse rate of dry atmospheres}
\author{Emilio Alvarez Navarro\inst{1} \and Bogar D\'iaz\inst{2,3,4} \thanks{\emph{e-mail:} bdiaz@iem.cfmac.csic.es} 
\and Miguel \'Angel García-Ariza\inst{3} \and J. E. Ramírez\inst{4,5}% etc
\thanks{\emph{e-mail:} jerc.fis@gmail.com }%
}                     % Do not remove
\institute{Departamento de Actuaria, F\'isica y Matem\'aticas, Universidad de las Am\'ericas Puebla, Santa Catarina Martir, 72820 Puebla, M\'exico \and Instituto de Estructura de la Materia, CSIC, Serrano 123, 28006 Madrid, Espa\~na \and Departamento de Matem\'aticas, Instituto de Ciencias, Benem\'erita Universidad Aut\'onoma de Puebla, Ciudad Universitaria, 72570, Puebla, Puebla, M\'exico  \and Facultad de Ciencias F\'isico Matem\'aticas, Benem\'erita Universidad Aut\'onoma de Puebla,
Apartado Postal 165, 72000 Puebla, Puebla, M\'exico \and Departamento de F\'isica de Part\'iculas, Universidad de Santiago de Compostela, E-15782 Santiago de Compostela, Espa\~na}
\date{}

%\maketitle
\abstract{
We study the effect of the second virial coefficient on the adiabatic lapse rate of a dry atmosphere. To this end, we compute the corresponding adiabatic curves, the internal energy, and the heat capacity, among other thermodynamic parameters. We apply these results to Earth, Mars, Venus, Titan, and the exoplanet G1 851d, considering three physically relevant virial coefficients in each case: the hard-sphere, van der Waals, and the square-well potential. These examples illustrate under which atmospheric conditions the effect of the second virial coefficient is relevant.
Taking the latter into account yields corrections towards the experimental values of the lapse rates of Venus and Titan in some instances.}
\maketitle
\PACS{Lapse rate}
%      {PACS-key}{discribing text of that key}   \and
%      {PACS-key}{discribing text of that key}
 % end of PACS codes
%end of abstract
%

%\maketitle

\section{Introduction}
The \textit{lapse rate} of the atmosphere of a planet, which we denote by $\Gamma$, is the rate of change of its temperature with respect to height. Its experimental value in some astronomic objects within the solar system is well known \cite{Kasprzak1990,Lindal1983,Mokhov2006,haberle}. However, reproducing this number theoretically becomes a rather difficult task, since atmospheres comprise many elements, contain traces of vapor, and undergo many thermodynamic processes \cite{Folkins2002,Catling2015,Stevenson}.

In the usual theoretical approach to $\Gamma$, one considers that each \textit{parcel} of atmosphere is in thermodynamic equilibrium, and rises up exchanging no heat with its surroundings. This framework is known as \textit{adiabatic atmosphere}, and the corresponding $\Gamma$ is called \textit{adiabatic lapse rate}. The latter is computed by means of the hydrostatic equation and an equation of state that describes the gas in the parcels of atmosphere. The simplest choice, the equation of an ideal gas, yields a lapse rate whose value is far from experimental data \cite{Stone1975,Catling2015}. A means to improve this result is \textit{correcting} the equation of state using the virial expansion, like we do in this work. 

As is well known, it is possible to derive the virial equation from the principles of statistical mechanics, relating the equation of state of a gas to the forces between molecules \cite{mcquarrie2000}. We are particularly interested in the second virial coefficient, which represents the contribution to the corresponding equation of state due to collisions occurring in clusters of two molecules. Its experimental values for different substances are known \cite{Bienkowski}.

An alternative path to overcoming the problem of choosing the right equation of state for an adiabatic atmosphere is followed in ref. \cite{Staley1970}, for the particular case of the atmosphere of Venus. Therein, a general expression for the lapse rate of real gases is derived, in terms of the compressibility factor $Z$ and the isobaric specific heat capacity $c_p$. This allows for the use of tabulated data of the former, instead of resorting to a theoretical or semi-empirical equation of state to compute $\Gamma$. In spite of yielding a value in good agreement with the experimental result, a shortcoming of this approach is the availability of tabulated data for other planets. Furthermore, the use of the latter renders it impossible to determine the origin of the corrections to $\Gamma$. Namely, we cannot determine how each virial coefficient and the vibration of the molecules contribute to the lapse rate. In contrast, the aim of our work is precisely to figure out the role of the second virial coefficient in the computation of $\Gamma$, as we explained before.

It is worth mentioning that, as we are only interested in the effect of the second virial coefficient, the resulting lapse rate shall be far from  experimental data in particular instances, even when we observe differences with respect to the ideal gas prediction. The reason is that in these examples other effects become relevant. For example, it is well known that latent heat release in the atmosphere of the Earth has a much larger effect on the adiabatic lapse rate than the virial coefficients and molecular vibrations. This renders the so-called \textit{moist lapse rate} approach a rather suitable one \cite{fleagle,holton,Catling2015,blundell2009}.

In order to study the contribution of the second virial coefficient, denoted henceforth by $B(T)$, to the adiabatic lapse rate, we determine the correction of the adiabatic curves and the hydrostatic equation arising therefrom. We apply the results to compute the dry lapse rate of the Earth, Mars, Venus, Titan, and the exoplanet G1 851d, considering that the corresponding atmospheres are monocomponent, formed up by the most abundant gas (see table~\ref{tab:table1}). In each case, we analyze three instances of second virial coefficients, related to simple fluid models that include the presence of molecular interactions. Namely, the hard-sphere, van der Waals, and the square-well potential. Furthermore, we consider no contributions due to molecular vibrations, whence we take the heat capacity at constant volume to be $C_V^\text{IG}=5R/2$ for linear molecules, according to the classical equipartition theorem, as is the case of diatomic molecules and carbon dioxide \cite{mcquarrie2000,callen1985}.

We have organized this paper as follows. We briefly review the ideal gas approach to the adiabatic atmosphere in sect. \ref{model}. Next, in sect. \ref{sec:CIG}, we introduce the equation of state that we shall consider to compute the lapse rate: the virial expansion up to second order of a simple fluid. Then, we compute the \textit{point lapse rate} (this is, the lapse rate at each parcel of atmosphere) for dry atmospheres.  In sect. \ref{sec:Resu}, we  apply our results to Earth, Mars, Venus, Titan, and the exoplanet G1 851d, considering the aforementioned models for $B(T)$. Finally, section \ref{sec:CR} is devoted to closing remarks and perspectives.

\section{The adiabatic lapse rate of an ideal gas atmosphere}\label{model}

In this sect. we reproduce the well-known derivation of the adiabatic lapse rate for a dry atmosphere formed by an ideal gas (see, for instance, refs. \cite{Stone1975,Catling2015,Vallero2014}). We shall follow an analogous procedure to obtain the adiabatic lapse rate for the viral expansion up to second order in sect. \ref{sec:CIG}.

It is common knowledge that the pressure $P$ of air at any height in the atmosphere decreases with height at a rate given by the \textit{hydrostatic equation}:
\begin{equation}
\de P=-\rho g \de z,
\label{phidro}
\end{equation}
\noindent where $\rho$ denotes the mass density of the gas and $g$ is the magnitude of acceleration due to gravity close to the surface of a planet. 

In the case of ideal gases, $P=nk_\text{B} T$, with $n$ denoting the particle number density, $T$  the temperature of the gas, and $k_\text{B}$ Boltzmann's constant. Taking into account that $\rho$ is expressed in terms of the molecular mass $m$ as $\rho=nm$, we can rewrite eq. \eqref{phidro} for ideal gases as

\begin{equation}
T\frac{\de P}{P}=-\frac{mg}{k_\text{B}}\de z.
\label{phidro2}
\end{equation}

The left-hand side of the equation above contains two independent state variables, temperature, and pressure. Therefore, in order to solve it, we require to relate somehow $P$ and $T$ to each other. This is done by considering different thermodynamic processes. For instance, a straightforward ---but rather unrealistic--- approach consists of considering $T$ to be constant, which is called \textit{isothermal atmosphere}. 

A usual and more realistic approach \footnote{It is worth mentioning that despite being a more accurate approach, the model of adiabatic atmosphere is only valid for the troposphere.}, called \textit{adiabatic atmosphere}, assumes that each \textit{parcel} of atmosphere rises exchanging no heat with its surroundings. The states that they may attain are thus constrained to lie on curves representing adiabatic processes. For ideal gases, adiabatic processes satisfy $P^{1-\gamma}T^\gamma=\text{const.}$, where $\gamma$ denotes, as usual, the heat capacity ratio $C_P/C_V$. Substituting this information in eq. \eqref{phidro2}, we have that

\begin{equation}\label{eq:hydro1}
\frac{\de T}{\de z}=-\left( \frac{\gamma-1}{\gamma} \right) \frac{mg}{k_\text{B}}.
\end{equation}

We rewrite eq. \eqref{eq:hydro1} in terms of physically measurable quantities. Recall that the heat capacity ratio is related to the molar mass $M_\text{mol}$ of a gas by $(\gamma-1)/\gamma=M_\text{mol}k_\text{B}/(mC_P)$. Thus, equation \eqref{eq:hydro1} takes the form

\begin{equation}
\frac{\de T}{\de z}=-\frac{M_\text{mol} g}{C_P}.
\label{Tz}
\end{equation}

According to eq. \eqref{Tz}, temperature decreases linearly with height. The corresponding proportionality constant, $M_\text{mol} g/C_P$, is called \textit{adiabatic lapse rate} \cite{Catling2015,blundell2009}. 

By means of eq. \eqref{Tz}, we may compute the value of the lapse rate on any astronomical object having an ideal gas atmosphere. Those corresponding to the astronomical objects of our interest are compared to experimental data \cite{Catling2015,Stone1975,Hummel1981,McKay1997} in table \ref{tab:table1}. We suppose that atmospheres are composed only of the most abundant gas in every case. It is important to remark that, under this assumption, the ideal gas prediction in Venus differs from the one reported in the literature. The reason is that commonly ---and erroneously---, the $C_P$ used in eq. \eqref{Tz} corresponds to real gases \cite{Staley1970,hilsenrat}. %(see also eq. \eqref{CpCv}, below). 
Similarly, the prediction on Earth presented here is also different from the usual one of $\SI{9.8}{\kelvin\per\kilo\meter}$ \cite{Catling2015} (notice that the former is closer to the experimental value).

\begin{table}[ht!] \centering
    \caption{Most abundant gas in each atmosphere, values of the lapse rate predicted by the ideal gas ($\Gamma_\text{IG}$), the corresponding experimental measurement ($\Gamma_\text{Exp}$), and atmospheric conditions on the surface of each astronomical object.}
    \label{tab:table1}
    
    \begin{tabular}{|P{1.9cm}|P{1.6cm}|c|P{1.5cm}|P{1.5cm}|P{1.5cm}|P{1.5cm}| P{1.5cm}|}
 \hline 
      Astronomical object &	Major constituent &	Composition	& $\Gamma_{\text{IG}}$ [$\SI{}{\kelvin\per\kilo\meter}$]&	$\Gamma_{\text{Exp}}$ [$\SI{}{\kelvin\per\kilo\meter}$] & $P_0$ [$\SI{}{\kilo\pascal}$] &$T_0$ [$\SI{}{\kelvin}$]& $g$ [$\SI{}{\meter\per\second^2}$] \\
\hline
Earth &	N$_2$	& 78\%	& 9.44	& 6.5  &	101&	288& 9.80 \\
Titan &	N$_2$	& 94.2\%	& 1.30 &	1.38 &	150&	93.7& 1.35\\
Mars & CO$_2$ & 96\% & 5.61 & 2.5 &0.6	&215& 3.71\\
Venus &	CO$_2$	& 96.5\%	& 13.42	& 8.4 &	9200&	737& 8.87\\
Gl 581d & CO$_2$ & 96\%  & 30.70 & ---&---&---&20.30\\
\hline
    \end{tabular}
\end{table}

\section{Corrected ideal gases}\label{sec:CIG}

The equation of state of an ideal gas may be corrected by considering the interaction among its microscopic constituents. One way to achieve this is through the virial expansion. Up to second order it is given by \cite{Boltachev2006,Masters2008}
\begin{equation}
\frac{m^2P}{k_\text{B}T}=m\rho+B(T)\rho^2\frac{k_\text{B}}{R}.
\label{den2}
\end{equation}
The importance of the second virial coefficient $B(T)$ in the equation above lies on the fact that it represents pairwise interactions of molecules, and it can be experimentally measured \cite{wisniak,dymond2002}. Every model of a fluid has associated to it a particular form of  $B(T)$ containing free parameters whose values are adjusted to experimental data. We consider three of the most usual ones: the hard-sphere, van der Waals, and the square-well potential (see table~\ref{tab:table2}). Notice that when $B(T)\to0$ in eq. \eqref{den2} we recover the ideal gas equation. The latter is to be understood as the limit of the free parameters contained in $B(T)$ approaching to 0 (for all values of $T$) \footnote{This must not be confused with the Boyle temperature, which is the value of $T$ satisfying $B(T)=0$.}. In the models presented here, this limit is equivalent to vanishing the volume of the molecules comprising the fluid and \textit{turning off} interactions between them. We call this the {\it ideal gas limit}.

\subsection{Correction to the hydrostatic equation of an ideal gas and  modified lapse rate}

In order to incorporate $B(T)$ in the hydrostatic equation, we require an expression for $\rho$ in terms of the virial coefficient. This is obtained through eq. \eqref{den2}. Since it is a quadratic equation on $\rho$, we have in principle two solutions. However, one of them may be discarded on physical grounds: it yields negative values of $\rho$ in the ideal gas limit. The physically meaningful solution is thus given by
\begin{equation}\label{eq:rho}
\rho=\frac{mR}{2B(T)k_\text{B}} \left( \sqrt{1+\frac{4P}{RT}B(T)} -1 \right).
\end{equation}
\noindent Observe that the equation of state of an ideal gas is recovered in the corresponding limit.

Substituting eq. \eqref{eq:rho} in eq. \eqref{phidro} yields 

\begin{equation}
\de P=-\frac{M_\text{mol}g}{2B(T)} \left( \sqrt{1+\frac{4P}{RT}B(T)} -1 \right)\de z,
\label{dpgr}
\end{equation}
\noindent which is the hydrostatic equation for an atmosphere with parcels described by a virial expansion up to second order. As in sect. \ref{model}, we need to determine the adiabatic curves in order to calculate $\de P$.
%and then arrive at expression to the lapse rate as a function of the temperature only. We can give a formula without having the explicit form of the curves, by using that the adiabatic curves are defined by some function, $F(T,P)=0$, and as we see below, 
In the cases we consider (see subsect. \ref{CAC}), we can write $P=P(T)$ on each adiabatic curve, whence $\de P= P'\left(T\right) \de T$, with $P'\left(T\right)$ denoting the derivative of $P\left(T\right)$ with respect to $T$. Substituting this in eq. \eqref{dpgr} we have 
\begin{equation}\label{newdpgr}
-\frac{\de T}{\de z}=  \frac{M_\text{mol}g}{2B\left(T\right) P'\left(T\right)} \left( \sqrt{1+\frac{4P \left(T\right)}{RT}B\left(T\right)} -1 \right).
\end{equation}
We call the right-hand side of eq.\eqref{newdpgr} the \textit{point lapse rate} of an atmosphere. As expected, eq. \eqref{newdpgr} reduces to eq. \eqref{Tz} (ideal gas case) in the corresponding limit. Indeed, close to the ideal gas limit, the right-hand side of eq. \eqref{newdpgr} can be written as
\begin{eqnarray}\label{newlim2}
 - M_\text{mol}g  \frac{P\left(T\right)}{P'\left(T\right)RT} + \mathcal{O}\left( B\left( T \right) \right),
\end{eqnarray}
where $\mathcal{O}\left( B\left( T \right) \right)$ is a polynomial. In the ideal gas limit, $\mathcal{O}\left( B\left( T \right) \right)=0$, $P'\left(T\right)=\gamma P(T)/(\gamma-1)T$ (see sect. \ref{model}), with $\gamma/(\gamma-1)=C_P/R$, which yields eq. \eqref{Tz}. We obtain $P=P\left(T\right)$ for the three models under consideration in the following subsect. \ref{CAC}.

\subsection{Correction to the adiabatic curves of an ideal gas}\label{CAC}

We shall use eq. \eqref{den2} to find the form of the corresponding adiabatic curves. To this end, it suffices to consider only one mole of gas, as will be done. In terms of the volume $V$ of the gas, the virial expansion up to second term yields \cite{Strandberg1977,matsumoto2010}

\begin{equation}
\frac{P}{RT}=\frac{1}{V}+\frac{B(T)}{V^2}.
\label{eqvirial}
\end{equation}

Recall that for an adiabatic process on one mole of gas, $\de U=-P\de V$. We compute $\de U$ using the caloric eq. $\de U=C_V\de T+(\partial U/\partial V)_T\de V$, which is corrected by eq. \eqref{eqvirial}. Indeed, taking into account the Maxwell relation $\left(\partial U/\partial V  \right)_T=T^2 \left( \partial \left( P/T  \right)/\partial T  \right)_V$, we have that

\begin{equation}\label{eq:calorica}
\de U=C_V\de T+\frac{R}{V^2}T^2B'(T)\de V,
\end{equation}
where $B'(T)$ is the derivative of $B(T)$ with respect to $T$.

The corrected equation for the adiabatic curves in a $V$-$T$ diagram is then

\begin{equation}
\frac{C_V}{T}\de T +\frac{R}{V} \left( 1+\frac{TB'(T)+B(T)}{V}  \right)\de V=0.
\label{dca}
\end{equation}
The left-hand side of eq. \eqref{dca} is an exact differential $\de f$.
%with
%\begin{equation}
%f(V,T)=R\ln V+\int  \frac{ C_V}{ T} \de T.
%\label{fsol}
%\end{equation}
In order to obtain an explicit form of the $f$, we need to know the form of $C_V$, which is also corrected by eq. \eqref{eqvirial}. 
\subsection{Correction to the heat capacity and the adiabatic curves}\label{subsect:HCAC}
We shall now determine $C_V$ considering eq. \eqref{eqvirial}.
In order to achieve this, we observe that 
\begin{equation}\label{eq:eqdiffv}
\left( \frac{\partial}{\partial V} \left( \frac{1}{T}\right) \right)_U=\left( \frac{\partial}{\partial U} \left( \frac{P}{T}\right) \right)_V,
\end{equation}
\noindent which follows from the Second Law of Thermodynamics in the entropic representation. Upon substituting eq. \eqref{eqvirial} in eq. \eqref{eq:eqdiffv}, we obtain
\begin{equation}
\left( \frac{\partial T}{\partial V}   \right)_U= -RT^2\frac{B'(T)}{V^2} \left( \frac{\partial T}{\partial U}   \right)_V.
\label{edU}
\end{equation}
The latter has to be solved for each particular form of $B(T)$ to obtain the corresponding internal energy (and hence, $C_V$). In table \ref{tab:table2}, we show both $U$ and $C_V$ for the three different virial coefficients under consideration. We point out the fact that the solution of eq. \eqref{edU} is given by a family of functions. We have chosen the ones that reduce correctly to those corresponding to ideal gases in the appropriate limit. It is worth mentioning that our approach reproduces the results obtained from the point of view of statistical mechanics \cite{matsumoto2010}. 

\begin{figure}[ht!]
\centering
\includegraphics[scale=1.13]{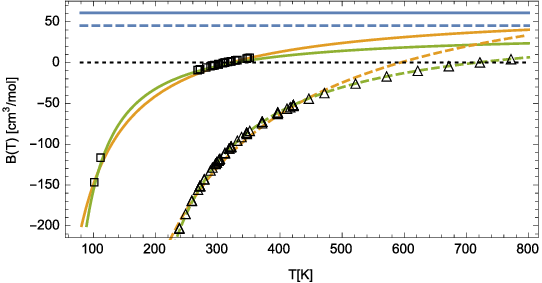}
\caption{Fits to experimental data ($\square$ and $\triangle$ for N\textsubscript{2} and CO\textsubscript{2}, respectively) of the second virial coefficients of N\textsubscript{2} (solid lines) and CO\textsubscript{2} (dashed lines) for the square well potential (green) and the van der Waals fluid (yellow). The values for the hard sphere (blue) interaction were calculated as in ref. \cite{Bienkowski}. The dotted line indicates $B(T)=0$.}
\label{fig:virial}
\end{figure}

For the three models under consideration, $U$ can be written as
\begin{equation}
U= C_V^\text{IG}T-\frac{RT^2}{V}B'(T). \label{C_V fina}
\end{equation} Using eq. \eqref{C_V fina} to compute $C_V$ and substituting in eq. \eqref{dca} yields
\begin{equation}\label{adibaticcurves}
 f(V,T)=  R\ln V+C_V^\text{IG}\ln T-\frac{R}{V}\mathcal{B}\left(T\right)=\text{const.},
\end{equation}
defining $\mathcal{B}(T)=TB'(T)+B(T)$. It is clear that in the ideal gas limit we recover the adiabatic curves of ideal gases. We can rewrite eq. \eqref{adibaticcurves} as
\begin{equation}
    V^{0.4}Te^{-0.4\mathcal{B}(T)/V}=\varepsilon_0,
\label{eq:CA-V}
\end{equation}
where we have used $C_V^{IG}=5R/2$. The constant $\varepsilon_0$ is determined by the atmospheric conditions on the surface (see table \eqref{tab:table1}). The free parameters of $B(T)$  (see tables \ref{tab:table2} and \ref{tab:pfits}) are fit to experimental data \cite{Dadson1967,Sevastyanov1986,dymond2002}, as is shown in fig. \ref{fig:virial}.
We solve eq. \eqref{eq:CA-V} for $V$, obtaining
\begin{equation}\label{eq:v(t)}
V(T)=\frac{\mathcal{B}(T)}{W\left( \mathcal{B}(T)  \left( \frac{T}{\varepsilon_0} \right)^{2.5} \right)},
\end{equation}
where $W$ is the Lambert function \footnote{The Lambert function is defined by the implicit equation $W(z) \exp W(z) = z$}. Notice that for the models under consideration we have $\mathcal{B}(T)>0$. Finally, if we substitute eq. \eqref{eq:v(t)} in eq. \eqref{eqvirial} we obtain the adiabatic curve $P=P(T)$. 

%\begin{table}[ht!]\centering
 %   \caption{Atmospheric conditions on the surface of each body.}
  %  \label{tab:table4}
   % \begin{tabular}{|c|c|c|}
% \hline
%Celestial object &	$P_0$ [$\SI{}{\kilo\pascal}$] &$T_0$ [$\SI{}{\kelvin}$] \\
%\hline
%Earth&	101&	288\\
%%Titan&	150&	93.7\\
%Mars&0.6	&215	\\
%%Venus&	9200&	737\\
%\%hline
 %   \end{tabular}
%\end{table}

\begin{table}[ht!]\centering
    \caption{Explicit forms of the second virial coefficient and their corresponding internal energy and specific heats at constant volume. We denote by $C_V^\text{IG}$ the heat capacity at constant volume of an ideal gas.}
    \label{tab:table2}
    \begin{tabular}{|c|c| c | c|}
 \hline
Model & $B(T)$ & $U$ &$C_V$ \\
\hline
van der Waals & $\displaystyle a+\frac{b}{T}$ & $\displaystyle C_V^{\text{IG}}T+\frac{bR}{V}$ & $C_V^{\text{IG}}$ \\
Square-Well & $\displaystyle a+ce^{b/T}$ & $\displaystyle C_V^{\text{IG}}T+\frac{cbR}{V}e^{b/T}$ & $\displaystyle C_V^{\text{IG}}-\frac{cb^2R}{T^2V}e^{b/T}$ \\
Hard Sphere & $a$ & $\displaystyle C_V^{\text{IG}}T$ & $C_V^{\text{IG}}$ \\
\hline
    \end{tabular}
\end{table}

\begin{table}[ht!]\centering
    \caption{Values of the parameters as a result of a fit to the experimental data for the second virial-coefficient.}
    \label{tab:pfits}
    \begin{tabular}{|c|c| c | c|}
 \hline
Model-Gas & $a$ & $b$ &$c$ \\
\hline
van der Waals N$_2$& $67.28\pm1.02$ $\SI{}{\cubic\centi\meter\per\mole}$&-21662.4$\pm$249.7 $\SI{}{\cubic\centi\meter \kelvin\per\mole}$ & --- \\
Square-Well N$_2$ &124.99$\pm$16.71 $\SI{}{\cubic\centi\meter\per\mole}$ &116.78$\pm$10.9 $\SI{}{\kelvin}$  & -87.64$\pm$14.53 $\SI{}{\cubic\centi\meter\per\mole}$ \\
van der Waals CO$_2$& 130.76$\pm$5.07 $\SI{}{\cubic\centi\meter\kelvin\per\mole}$ &-77664.2$\pm$1631 $\SI{}{\cubic\centi\meter\per\mole}$ & --- \\
Square-Well CO$_2$ &140.49$\pm$1.73 $\SI{}{\cubic\centi\meter\per\mole}$  &322.14$\pm$2.22 $\SI{}{\kelvin}$ & -89.65$\pm$1.23 $\SI{}{\cubic\centi\meter\per\mole}$ \\
\hline
    \end{tabular}
\end{table}

\begin{figure}[ht!]
\centering
\includegraphics[scale=1]{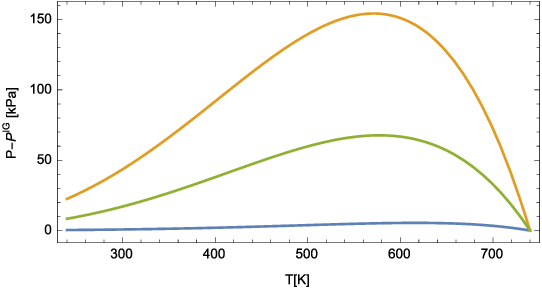}
\caption{Deviation of the adiabatic curves that includes the second virial coefficient  from the adiabatic curves of the ideal gas in Venus. Van der Waals (yellow), square-well (green) and hard sphere (blue).} %molecular potential interactions.}
\label{fig:adcurves}
\end{figure}

\begin{figure}[h]
\centering
\includegraphics[scale=1]{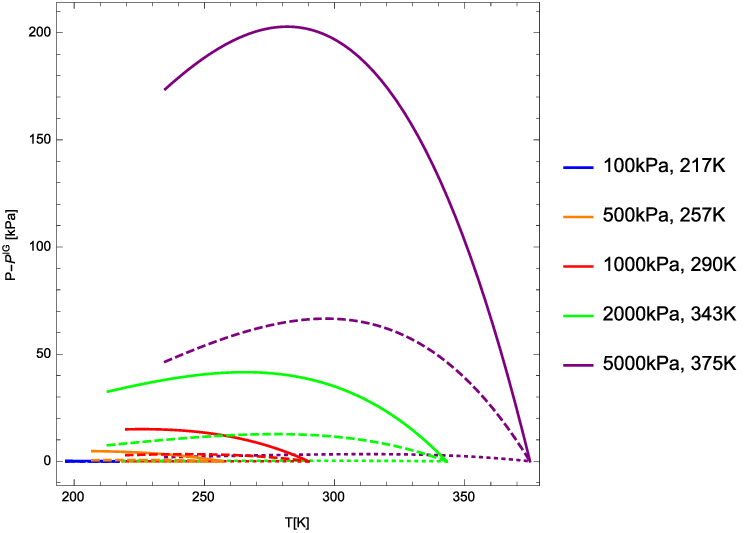}
\caption{Deviation of the adiabatic curves from the adiabatic curves of the ideal gas for the second virial coefficient of van der  Waals (solid lines), square-well (dashed lines) and hard sphere (dotted lines) in the exoplanet Gl 581d.}
\label{fig:adExo}
\end{figure}

%\textcolor{red}{The most significant correction to the adiabatic curves is found in Venus and the exoplanet Gl 581d, as can be seen in figs. \ref{fig:adcurves} and \ref{fig:adExo}. In the cases of Earth, Mars, and Titan, we do not appreciate any substantial difference. This is to be expected on Earth and Mars because of their particular atmospheric conditions. The corrections to pressure in the case of Titan are of the order of $\SI{1}{\kilo\pascal}$, which is negligible in the ranges of pressure close to the surface of the body.} 

In order to illustrate the deviation of the adiabatic curves from the ideal case, in figs. \ref{fig:adcurves} and \ref{fig:adExo} we show the cases of Venus and the exoplanet Gl 581d, respectively. Note that adiabatic curves approach the ones of an ideal gas as the troposphere reaches the STP conditions. In the cases of Earth, Mars, and Titan, the curves also deviate, but it is less than $\SI{1}{\kilo\pascal}$. This is expected on Earth and Mars because the gases in their tropospheres behave like ideal ones, as we explain in subsect. \ref{sub-cig} (see also table~\ref{tab:table1}).

In sect. \ref{sec:Resu} we apply eq. \eqref{newdpgr} to some astronomical objects. In order to compare the result this yields to the corresponding ideal gas predictions and experimental values, we integrate the latter numerically in a suitable interval and define $\Gamma$ to be $\Delta T/\Delta z$.

\section{Results}\label{sec:Resu}

In this sect. we compute the tropospheric lapse rate of some astronomical objects having available experimental data regarding their real atmospheric conditions. In order to enhance the exposition, we have divided our analysis in two groups: bodies with atmospheres close to ideal gases (Earth and Mars), and atmospheres in extreme conditions of pressure or temperature (Titan and Venus). We also discuss the possible values of the lapse rate of the exoplanet G1 851d under a range of atmospheric conditions where it could be habitable \cite{Hu}. It is worth mentioning that the real conditions for this exoplanet are unknown, and therefore we are unable to compare our results to any experimental value. However, we can compare our results to the ideal gas prediction, which illustrates when the second virial coefficient becomes relevant.

The values involved in all our calculations are listed in tables \ref{tab:table1}, \ref{tab:pfits}, and \ref{tab:table5}.

%table5
\begin{table}[ht!]\centering 
    \caption{Values of $\varepsilon_0$ (in $\SI{}{\meter^{1.2}\kelvin}$) for the adiabatic curves considering all virial coefficients in table \ref{tab:pfits}, computed using \eqref{eq:v(t)}, where $V_0$ is calculated by using $P_0$ and $T_0$ in the physically meaningful solution for $V$ of Eq. \eqref{eqvirial},
$V=\frac{TR}{2P}\left(1+\sqrt{1+\frac{4P}{RT}B(T)} \right)$%, 
%by evaluating in $P_0$, and $T_0$, 
(provided that the other solution is identically zero for ideal gases and is therefore discarded).} 
    \label{tab:table5}
    \begin{tabular}{|c|c|c|c|c|}
\hline
Model  &	Earth&	Titan&Mars &	Venus\\
\hline
van der Waals&	64.37&	11.27& 332.64 &	37.40\\
Square-Well&	64.40&	11.13& 332.63	&38.34\\
Hard Sphere&	64.46&	11.48& 332.66	&39.71\\
\hline
    \end{tabular}
\end{table}

\subsection{Astronomical objects with atmospheres close to ideal gas}\label{sub-cig}

\begin{figure}[ht!]
\centering
\includegraphics[scale=1]{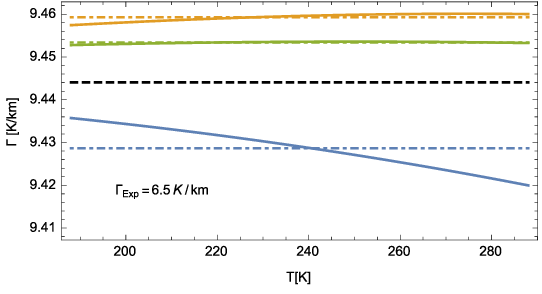}
\caption{Point lapse rate (solid curve) on Earth and average lapse rate (dash-dotted line) for the virial coefficients of van der Waals (yellow), square well (green), and hard spheres (blue). The dashed black line corresponds to the ideal gas prediction.} 
\label{fig:Tlr}
\end{figure} 

In this subsect. we present the value of the lapse rate of Earth and Mars that eq. \eqref{newdpgr} yields. Considering a dry lapse rate, their atmospheres behave like ideal gases due to their atmospheric conditions: Earth's is close to STP, and Mars' is diluted (which drives it far from the coexistence curve \cite{wisniak}). This is reflected in their modified lapse rates, as we shall see. The numerical integration is carried out in the temperature intervals $\SI{188}{\kelvin}<T<T_0$ for the Earth, and $\SI{130}{\kelvin}<T<T_0$ for Mars. The former is determined by the length of the troposphere ($\SI{10}{\kilo\meter}$), whereas the latter makes use of experimental data \cite{haberle}. 

In fig.~\ref{fig:Tlr} we show the results for the lapse rate of Earth. As was anticipated, the result is very close to the ideal gas prediction, regardless of the model under consideration. Similarly, we do not observe differences up to thousandths with respect to the ideal gas prediction ($\SI{5.612}{\kelvin/\kilo\meter}$) in the case of Mars. This is to be expected, as the range of pressure prevents the condensation or deposition of CO$_2$, whence molecular interactions become negligible. The difference with respect to the experimental value is due to additional heating coming from the absorption of solar radiation by suspended dust particles \cite{haberle}.

\subsection{Astronomical objects with atmospheres in extreme conditions of pressure or temperature}\label{sub:Ext}

There are two physical situations in which  molecular interactions become relevant: high temperatures and pressures, like in Venus, and states close to the boiling point but outside the liquid-gas coexistence curve (where latent heat is negligible), as is the case of Titan. We present our results for the lapse rate of these two objects in this subsect.

\begin{figure}[ht!]
\centering
\includegraphics[scale=1]{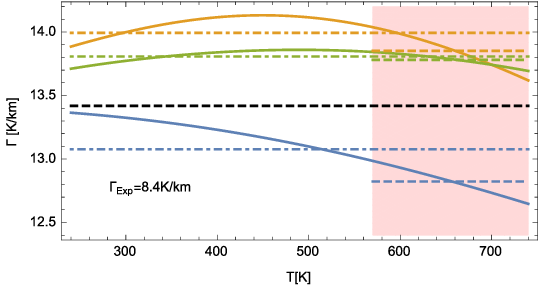}
\caption{Point lapse rate (solid curve) in Venus and average lapse rates in the whole troposphere (dash-dotted line) and near the surface (dashed line) for the virial coefficients of van der Waals (yellow), square well (green), and hard spheres (blue). The shaded region corresponds to the first $\SI{10}{\kilo\meter}$ of the atmosphere. The ideal gas prediction is represented by a black dashed line.}
\label{fig:Vlr}
\end{figure}

The lapse rate of Venus was computed in the interval $\SI{240}{\kelvin}<T<T_0$ (see table \ref{tab:table1}). The lower bound for $T$ corresponds to the length of the troposphere of Venus in the ideal gas approach ($\SI{45}{\kilo\meter}$). We observe a correction towards experimental data. However, because of its high temperature, molecular vibrations become relevant in this context. Therefore, the resulting lapse rate is still inaccurate. The corrected adiabatic curves are mostly deviated from those of ideal gases for temperatures above $\SI{570}{\kelvin}$  (corresponding to \textit{ca.} $\SI{20}{\kilo\meter}$, according to $\Gamma_\text{Exp}$), which is depicted in the shaded region of fig.~\ref{fig:Vlr}. As expected, there is a greater correction to $\Gamma$ herein. As pressure and temperature decrease higher up in the atmosphere, the corrected adiabatic curves resemble those of an ideal gas. As a result, the correction to the lapse rate in the whole of the troposphere is moved towards the ideal gas prediction.

\begin{figure}[ht!]
\centering
\includegraphics[scale=1]{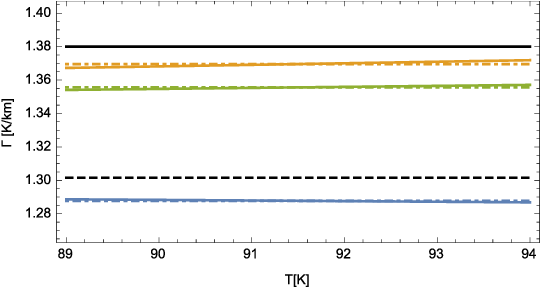}
\caption{Point lapse rate (solid curve) in Titan and average lapse rate (dash-dotted line) for the virial coefficients of van der Waals (yellow), square well (green), and hard spheres (blue). The black lines corresponds to the ideal gas prediction (dashed), and to the experimental value (solid).}
\label{fig:Tilr}
\end{figure}

On the other hand, the low temperatures of Titan render molecular vibrations negligible. These particular atmospheric conditions yield a value for the lapse rate very close to the experimental one. In this case, we considered only the first $\SI{3.5}{\kilo\meter}$ of the $\SI{55}{\kilo\meter}$-long troposphere due to data availability (see ref. \cite{Lindal1983}). The corresponding temperature range is $\SI{89}{\kelvin}<T<T_0$. It is important to point out that, even though the correction to the adiabatic curves is small, there is a noticeable shift towards the experimental value for potentials with interactions (see fig. \ref{fig:Tilr}).

\subsection{Exoplanet G1 581d}

The search for planets similar to Earth has received a lot of attention in recent years \footnote{An example of this is the Exoplanet Exploration Program of NASA, which aims are to characterize their properties and to identify which of them could harbor life, https://exoplanets.nasa.gov/ }. In 2007, the planet Gl 581d was detected orbiting the M-dwarf Gl 581 \cite{udry2007}. It has around eight times Earth's mass and gravity of $\SI{20.3}{\meter/\second^2}$. Its location renders it likely habitable \cite{Hu}. In order to possess liquid water, its atmosphere must consist of 96\% of CO$_2$ in a wide range of pressures and temperatures at the surface level \cite{Hu,wordsworth,vonParis}. The computation of the lapse rate that we present in this subsect. took into account these considerations, displayed in table~\ref{tab:exo}. The latter also contains the corresponding value of $\varepsilon_0$.

\begin{table}[ht!]\centering
    \caption{Atmospheric conditions on the surface of Gl 581d, and the corresponding value of $\varepsilon_0$.} 
    \label{tab:exo}
    \begin{tabular}{|c|c|c|c|c|c| }
\hline
$T_0$ [K]&	$P_0$ [kPa]&	Model&	$\varepsilon_0$ [$\SI{}{\meter^{1.2}\kelvin}$]\\
\hline
&	& van der Waals &	43.19\\
217 & 100 &	Square-Well&	42.97\\
& &		Hard Sphere&	43.54\\
\hline
&& van der Waals &	28.13\\
257& 500 &		Square-Well&	27.88\\
& &		Hard Sphere&	28.98\\
\hline
	&&	van der Waals&	24.79\\
290& 1000&		Square-Well&	24.65\\
& &		Hard Sphere&	26.01\\
\hline
	&&	van der Waals&	23.26\\
343 & 2000 &		Square-Well&	23.35\\
& &		Hard Sphere&	24.93\\
\hline
	&&	van der Waals&	16.69\\
375& 5000&		Square-Well&	17.06\\
& &		Hard Sphere&	19.57\\
\hline
    \end{tabular}
\end{table}

In this case, the temperature interval is bounded from below by conditions that allow for the coexistence of two or more phases (gas-liquid, gas-solid, or the triple point of CO$_2$), so that other effects (like latent heat) may be neglected. %\sout{In fig.~\ref{fig:adExo}, we show the adiabatic curves of each model and each atmospheric condition under consideration}.
Therefore, the adiabatic curves in fig.~\ref{fig:adExo} are considered in such a way that avoids the coexistence curve. Moreover, notice that when atmospheric conditions are close to STP, this deviation decreases, as we discussed in subsect.~\ref{subsect:HCAC}.

\begin{figure}[ht!]
\centering
\includegraphics[scale=1]{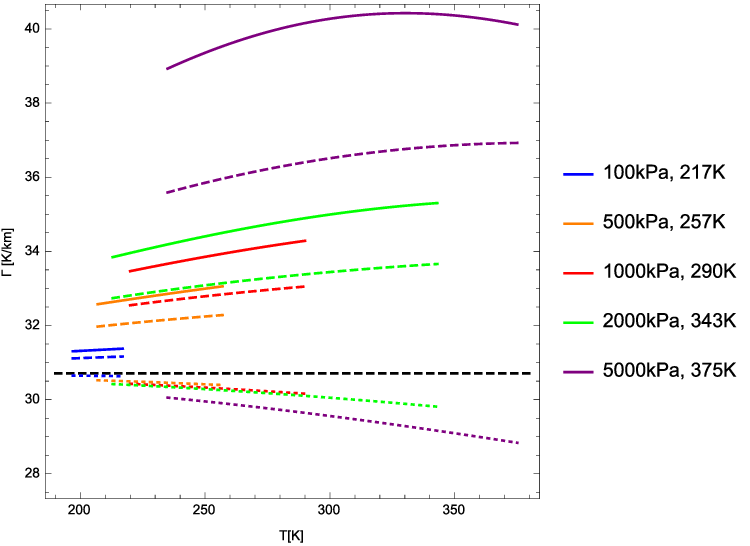}
\caption{Point lapse rate on Gl 581d  for the virial coefficients of van der Waals (solid lines), square well (dashed lines), and hard spheres (dotted lines). The ideal gas prediction is represented by a black dashed line.}
\label{fig:lrExo}
\end{figure}

Remarkably, this example shows the atmospheric conditions under which the effect of the second viral coefficient on the lapse rate is relevant as fig.~\ref{fig:lrExo} illustrates. Notice that for the lower values of temperature and pressure ($\SI{<500}{\kilo\pascal}$), the lapse rate of the van der Waals and square-well models overestimate the ideal gas prediction, in contrast to the hard-sphere model. This is in full agreement with the results obtained in Titan, which has similar atmospheric conditions (see subsect. \eqref{sub:Ext}). Furthermore, as expected, the value of the lapse rate approaches to the ideal gas prediction for low pressures.

\section{Concluding Remarks and Perspectives}\label{sec:CR}

We determined the effect of the second virial coefficient on the lapse rate of dry atmospheres by means of eq. \eqref{newdpgr}.The latter is a version of eq. (4), modified by the introduction of $B(T)$. The adiabatic curves were correspondingly modified. Their deviation from the ideal case depends on the atmospheric conditions. This is remarkably illustrated in Venus. Integrating eq. \eqref{newdpgr} numerically allowed us to identify $\Gamma$ as $\Delta T/\Delta z$. It is important to point out that all the equations that are corrected by $B(T)$ are validated by ideal gas limit.

In order to determine quantitatively the contribution of $B(T)$ to the lapse rate, we define
\begin{equation}\label{eta}
\eta=\frac{\Gamma_\text{IG}-\Gamma}{\Gamma_\text{IG}-\Gamma_\text{Exp}},
\end{equation}
Observe that when $0<\eta<1$, the corresponding values of $\Gamma$ are closer to experimental results than those predicted by the ideal gas model.

\begin{figure}[ht!]
\centering
\includegraphics[scale=1.]{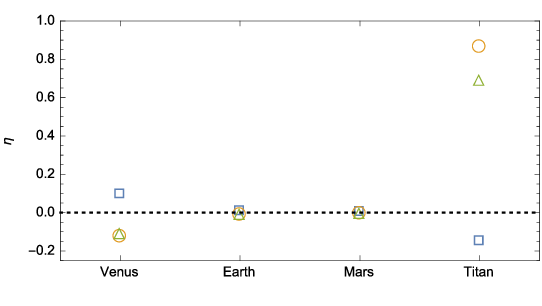}
\caption{Correction factor for the virial coefficients van der Waals ($\circ$), square well ($\triangle$), and hard spheres ($\square$), for the atmospheres of the four bodies under consideration. The dotted line indicates $\eta=0$.}
\label{fig:correction}
\end{figure}

We show the values of $\eta$ corresponding to the three models under consideration, for the four astronomical objects with available experimental data in fig.~\ref{fig:correction}.

When a gas behaves almost like an ideal one, the values of $\eta$ are close to 0 regardless of the model, as expected. In contrast, $B(T)$ becomes relevant in the extreme conditions of Titan and Venus. The nitrogen in the atmosphere of the former is close to its boiling point (\textit{ca.} $\SI{77}{\kelvin}$ at $\SI{101}{\kilo\pascal}$). Thus, molecules are close to each other, whence interactions between them are more frequent. This explains the values of $\eta$ corresponding to potentials that consider molecular interactions, viz., the square-well and van der Waals models. It is worth pointing out that these results improve previous ones (\textit{cf}. ref. \cite{Lindal1983}). 
On the other hand, under the conditions of Venus' atmosphere, carbon dioxide is far from its boiling point, which renders intermolecular interactions less relevant when compared to the kinetic energy within the gas.

Notice that when the temperature increases enough, $B(T)$ becomes positive. In that case, the first term is dominant, meaning that the interaction through elastic collisions is the most relevant.
Therefore, we might expect that the hard sphere model contributes most significantly to $\Gamma$, as was indeed the case of Venus. Up to second order in the virial expansion, a more accurate approach to the dry adiabatic lapse rate of Venus (and of the Earth) requires incorporating molecular vibrations. This can be done by making use of either tabulated experimental values or phenomenological equations for $C_P$ ( see, for instance, ref. \cite{shomate1954}). We do not expect an improvement in the lapse rate of Titan from this source, since the difference between the corresponding phenomenological $C_P$ and $7R/2$ is less than $\SI{0.02}{\joule\per{\mol\kelvin}}$. Rather, higher order contributions from the virial expansion might turn significant in this case.

Finally, taking into account gas mixtures makes the second virial coefficient (and all its derived quantities) dependent on both temperature and composition. Incorporating the latter to the framework presented herein might only be meaningful on the Earth, due to the predominance of one gas in the other atmospheres. However, under STP conditions, the corrections from this source will still yield values close to the ideal gas prediction. As we have mentioned before, the solution to this problem is provided by the moist lapse rate approach.

\section*{Acknowledgments}
Bogar Díaz and J. E. Ramírez are supported by a CONACYT postdoctoral fellowship. Miguel \'Angel García-Ariza is supported by a VIEP-BUAP postdoctoral fellowship.  The plots were carried out using Wolfram Mathematica 10${}^{\textrm{TM}}$.

%\bibliographystyle{EPJ}
%\bibliography{bib}

%
% BibTeX users please use
% \bibliographystyle{}
% \bibliography{}
%\begin{thebibliography}{}
%Author, \textit{Book title} (Publisher, place year) page numbers
% etc
%\end{thebibliography}

\end{document}